# Diodelike asymmetric transmission in hybrid plasmonic waveguides via breaking polarization symmetry

Heran Zhang[1,3,#], Fengchun Zhang[3,#], Yao Liang[2,#], Xu-Guang Huang[1,3,*], Baohua Jia[2,*]

[1]*Guangzhou Key Laboratory for Special Fiber Photonic Devices and Applications, South China Normal University, Guangzhou, 510006, China.*
[2]*Centre for Micro-Photonics, Faculty of Science, Engineering and Technology, Swinburne University of Technology, Hawthorn, Victoria 3122, Australia.*
[3]*Guangdong Provincial Key Laboratory of Nanophotonic Functional Materials and Devices, South China Normal University, Guangzhou, 510006, China.*
[#]*These author contribute equally*

*Corresponding authors: huangxg@scnu.edu.cn, bjia@swin.edu.au

# Abstract

The ability to control the asymmetric propagation of light in nanophotonic waveguides is of fundamental importance for optical communications and on-chip signal processing. However, in most studies so far, the design of such structures has been based on asymmetric mode conversion where multi-mode waveguides are involved. Here we propose a hybrid plasmonic structure that performs optical diode behavior via breaking polarization symmetry in single mode waveguides. The exploited physical mechanism is based on the combination of polarization rotation and polarization selection. The whole device is ultra-compact with a footprint of 2.95 × 14.18 μm$^2$, whose dimension is much smaller than the device previously proposed for the similar function. The extinction ratio is greater than 11.8 dB for both forward and backward propagation at λ = 1550 nm (19.43 dB for forward propagation and 11.8 dB for the backward one). The operation bandwidth of the device is as great as 70 nm (form 1510 to 1580 nm) for extinction > 10 dB. These results may find important applications in the integrated devices where polarization handling or unidirectional propagation is required.

## 1. Introduction

In 2008, Zhang *et al*. proposed the concept of hybrid plasmonic waveguides (HPWs) [1]. Since then, this topic has attracted considerable interest as promising candidate for integrated photonic circuits (IPCs) [1-4]. The tipical HPW structure consists of a dielectric nanowire and a metal strip, which are separated by a low-index dielectric material [1,5]. Due to their unique characteristics in subwavelength optical confinement [1], significant birefringence effect [2], polarization rotation [5], and relative long range propagation [3], HPWs can push conventional optical components to the subwavelength scale [5-7]. By manipulating the phase and polarization of the incoming light, the HPWs can control the propagation of light in unprecedented ways, enabling many applications, i.e., optical telecommunication and interconnects [4], polarization convter [5], nanofocusing [6] and one-way mode conversion [7].

The significant interest in fast, compact, and efficient optical computing and communications has led to the development of optical systems that perform one-way transmission or optical diode behavior. Conventionally, this is achieved by mainly using magneto-optic materials that can break the Lorentz symmetry, i.e., Faraday rotation [8] However, this technology suffers from the difficulty of integration, because the magneto-optic effect is extremely weak and it requires bulk optics to achieve considerate polarization rotation. Another challenge of this technology is the mechanical complexity, since other moving bulk optical components, such as polarizers, mirrors and wave-plates, are required [9].

For many practical applications, it is more desired to use compact, functional devices based on IPCs that are compatible with CMOS fabrication. For this reason, tremendous efforts have been devoted to create on-chip optical diodes, which can be generally divided into two main categories. The first type is non-reciprocal method based on spatiotemporal modulations [10,11] and Kerr non-linearities [12]. However, those designs are relatively large in size, the dimension of which can be anywhere from several tens of microns to several centimeters. Moreover, they require external modulations (strong light beam or dynamic electric field) as well, which is often undesired. The other category is to use reciprocal methods, different with breaking the reciprocity of light to realize optical diodes. Various types of reciprocal diodes have been demonstrated recently based on sonic crystals [13,14], HPWs [15] and photonic crystals [16-18]. Those devices offer asymmetric transmission mainly based on spatial symmetry breaking (asymmetric mode conversion). For example, they can convert an even fundamental mode to a higher order odd mode in the forward direction while blocking the propagation of the fundamental mode in the backward direction [16]. This, however, requires multimode waveguides. For on-chip signal processing, single mode waveguides are often more desired, because they have smaller footprint, lower propagation loss, and higher tolerance of waveguide roughness compared with the multimode ones. Consequently, spatial-symmetry-breaking schemes face a fundamental limitation for single mode waveguides where higher order modes are suppressed. Polarization is another degree of light that is orthogonal to spatial modes. Although spatial-symmetry-breaking schemes have been widely demonstrated for optical diodes, the corresponding demonstrations based on polarization-symmetry-breaking, especially with single mode HPWs, have been elusive, only having been reported in the context of chiral metamaterials [19].

In this work, we report optical diode behavior in a novel hybrid plasmonic structure that breaks polarization symmetry. A quasi-TM polarized mode (the dominant polarization component $\mathbf{E_T}$ points along the x-axis) is converted into another quasi-TE polarized mode ($\mathbf{E_T}$ points along the y-axis) in the forward propagation (from port C to port A). While the same polarized mode cannot travel back in the counter-propagation with the energy leaking away to another port B (Figure 1a).

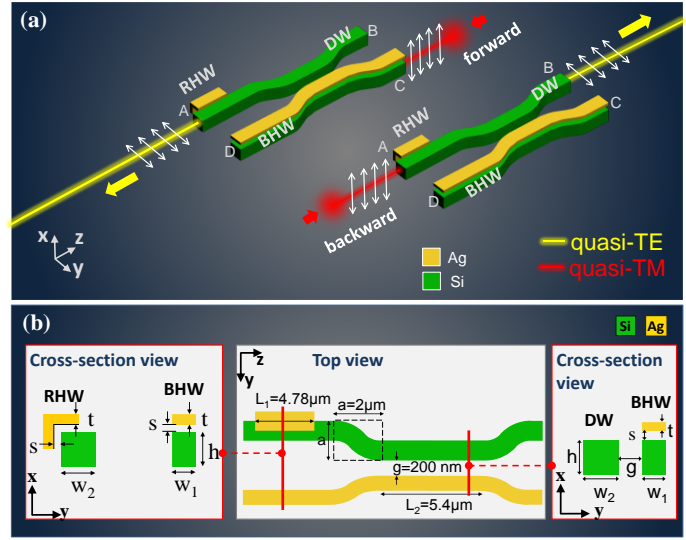

Figure 1. (a) Schematic illustration of the hybrid plasmonic waveguides system. An input quasi-TM mode can travel from port C to port A, but not vice versa. Also shown is the coordinate system used. (b) The inserted figure describes the top view and cross-sectional view of the hybrid plasmonic structures. The geometric details of the structure are ($w_1$, $w_2$, h, t, s) = (0.26, 0.34, 0.34, 0.1, 0.05) μm and ($L_1$, $L_2$, a, g) = (4.78, 5.4, 2, 0.2) μm. The whole structure is surrounded by silica, which is not shown for clarity.

## 2. Functionality and geometry details

Figure 1a shows the schematic configuration and function of the proposed structure. This includes a half-wave plate section and a directional coupler section, which respectively function as a polarization rotator and a polarization beam splitter. Three kinds of nanophotonic waveguides are involved in the hybrid system. A dielectric waveguide (DW) whose height and width are equal h = $w_2$ (Figure 1b), and a birefringent hybrid waveguide (BHW) with a metal (silver, Ag) cap on the top. They together make a polarization beam splitter, where 4 S-bend structures are used to couple and decouple light. The third kind is a rotating hybrid waveguide (RHW) with an L-shaped metal (silver, Ag) on the left top of the Si core, which works as a half-wave plate. The whole structure is surrounded by silica ($SiO_2$) and the operating wavelength is λ = 1.55 μm.

There are four physical ports (A, B, C and D) in the proposed device, and each of them support both quasi-TE and –TM polarized modes. For the diodelike asymmetric transmission, we only use three ports, A, B and C. When light travels from port C to port A (forward direction), a fundamental quasi-TM mode will first couple from the BHW to the adjacent DW and then get converted into a fundamental quasi-TE mode by the RHW at the output port A. However, in the counter propagation (A to C) (backward direction), a fundamental quasi-TM mode cannot travel back to the C port but instead, it gets converted into a quasi-TE mode and goes to output port B, since the newly converted quasi-TE mode cannot couples into the BHW due to phase-mismatch, which will be discussed later.

Figure 1b shows the detailed geometry parameters of the device. For the rotating hybrid waveguide, the width ($w_2$) and height (h) of the Si waveguide are equal, h = $w_2$ = 0.34 μm; the thickness and of L-shaped Ag cap (t) is t = 0.1μm, and the spacer (s) between L-shape Ag cap and silicon waveguide is s = 0.05 μm, and the whole polarization rotator length ($L_1$) is 4.78μm. For the polarization beam splitter part, the width ($w_2$) and height (h) of the DW are $w_2$ = h = 0.34 μm, while the width ($w_1$) and height (h) of the lower Si waveguide are 0.26 μm and 0.34 μm respectively; the thickness (t) of the metallic Ag cap t = 0.1μm, the spacer (s) between Ag cap and the Si waveguide s = 0.05 μm, the gap (g) between two adjacent waveguides g = 0.2 μm, and the whole polarization beam splitter part length $L_2$ = 5.4μm. The size of each S-bend is 2*2 μm$^2$ (the S-bend structures are taken from the structure database of Lumerical FDTD Solutions). The whole device is ultra-compact with a footprint of 2.95× 14.18 μm$^2$. The dielectric constants (ε) of the materials (Ag, Si, SiO$_2$) we used are $\varepsilon_{Ag}$ = -116.843+11.6819i, $\varepsilon_{Si}$ =12.0852, $\varepsilon_{SiO2}$ =2.08514, respectively.

## 3. Polarization rotator

We will first discuss the RHW and then the BHW and DW. The geometric details of the RHW is displayed in Fig. 2a while Fig. 2b shows the corresponding eigenmodes, whose dominant polarization components polarized at the angle φ = ±π/4, where φ is the angle down from the positive x-axis. The eigenmodes are obtained via the Eigenmode Solver that is available in the finite-different time-dominate (FDTD) Solutions package from Lumerical Inc.

As an example of non-paraxial light, the electric field of light in nanophotonic waveguides is not purely transverse. Unlike the plane wave in free space, where every point of the wavefront shares the same polarization state, the polarization of each point in nanophotonic waveguides is highly position-dependent. [21,22] However, the electric field of light at the central point of the Si DW is purely transversal regarding to a fundamental 0$^{th}$ order mode, where the amplitude of the dominant transverse polarization component ($E_T$) reaches its peak while the one of the longitudinal component ($E_z$) equals zero. Therefore, we use the center point polarization state to represent the predominant polarization state of a certain mode, and we use the word "quasi-" to make the description more accurate. Thus, it is reasonable to use the polarization state at the central point to represent the dominant polarization and phase of a certain fundamental mode [20].

Here, we choose the parameters as (h, $w_2$, t, s) = (0.34, 0.34, 0.1, 0.05) μm. Correspondingly, the effective indices for the two eigenmodes are respectively calculated to be $n_{π/4}$ = 2.47169 + 0.00352783$i$ and $n_{-π/4}$ = 2.30951 + 0.0020423$i$. When a quasi-TM mode ($E_{TM}$=exp[-i(wt-$n_{TM}k_0z$)]$e_{TM}$, where $e_{TM}$ is the unit vector) travels in the RHW, it will undergo a polarized modes coupling process. According to the polarized mode coupled theory [20], the dynamics of the output mode can be described by

$$\begin{cases} \dfrac{dE_{TE}}{dz} + in_{TE}k_0E_{TE} - i\kappa E_{TM} = 0 \\ \dfrac{dE_{TM}}{dz} + in_{TM}k_0E_{TE} - i\kappa E_{TE} = 0 \end{cases} \quad (1)$$

where $E_{TE/TM}$ represents the amplitude of two orthogonal modes at the output port, and κ = π/(2$L_C$) is the coupling coefficient with the coupling length $L_C$ = π/{[Re($n_{π/4}$) - Re($n_{-π/4}$)]*$k_0$}, $k_0$ = 2π/λ the free space wave number. For the output modes energies ( $P \propto |E|^2$ ), they can be written as,

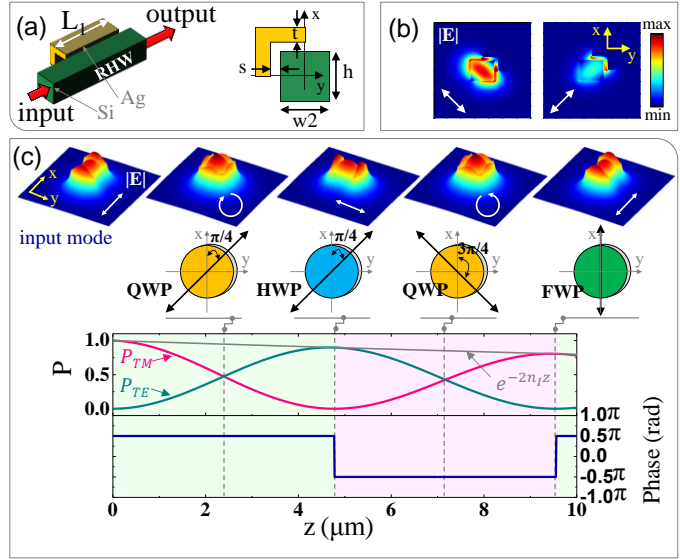

Figure 2. (a) The geometric details of the proposed RHW. ($w_2$, h, t, s) = (0.34, 0.34, 0.1, 0.05) μm. The whole structure is surrounded by SiO$_2$, which is not shown for clarity. (b) The eigenmodes supported in the RHW. (c) The theoretical dependence of the output modes powers ($P_{TM/TE}$) and relative phase ($\Delta\phi$) on the device length (z). The inserted figures show the corresponding wave plates and input and output modes. The white arrows indicate the polarization at the center point of Si core while the black arrows indicate the equivalent optical axis of various wave plates. Polarization evolution in the L-shaped hybrid waveguide output modes with different propagation length.

$$\begin{cases} P_{TE}(z) = \left|-i\sin(\kappa z)\right|^2 e^{-2\bar{n}k_0 z} \\ P_{TM}(z) = \left|\cos(\kappa z)\right|^2 e^{-2\bar{n}k_0 z} \end{cases} \quad (2)$$

provided that $P_{TE}(z=0) = 0$, $P_{TM}(z=0) = 1$, where z is the device length and $\bar{n} = [\text{Im}(n_{\pi/4}) + \text{Im}(n_{-\pi/4})]/2$.

The term $e^{-2\bar{n}k_0 z}$ denotes the propagation loss. In particular, the –i term in $P_{TE}(z)$ indicates that there is an intrinsic phase lag of π/2 between the output quasi-TE and -TM modes ($\Delta\phi = \phi_{TM} - \phi_{TE} = -\pi/2$). This characteristic is important, since it allows for the design of different kind of wave plates with different equivalent optical axis orientations by the choice of different device length (z), i.e., quarter-wave plates (QWPs), half-wave plates (HWPs) and full-wave plates (HWPs). For example, the choice of $z = 0.5L_C$ leads to a QWP with the equivalent optical axis orienting at φ = π/4 while a choice of $z = L_C$ results in a HWP with the equivalent optical axis pointing along φ = π/4, where φ is the angle down from the positive x-axis. To illustration this point, we plot the theoretical dependence of the output energies ($P_{TE/TM}$) and relative phase lag ($\Delta\phi$) on the device length (z) in Fig. 2c. In this work, we choose $z = L_C \approx 4.78$ μm, so that the RHW can convert an input quasi-TM mode into a quasi-TE mode, and *vice versa*.

## 4. Polarization beam splitter

Another section involved in our device is a polarization beam splitter, which consists of a BHW and a DW. Hybrid plasmonic waveguides can exhibit huge birefringence effect due to the asymmetry between the vertical and horizontal directions of structures. This effect can be steered to fabricate a polarization beam splitter. Fig. 3a shows the calculated results of the real part of effective indices of a BHW and a DW. In our calculation, the spacer between the Ag strip and the Si core for the BHW is s = 50 nm and the gap between two adjacent waveguides is g = 200 nm (Fig. 1b).

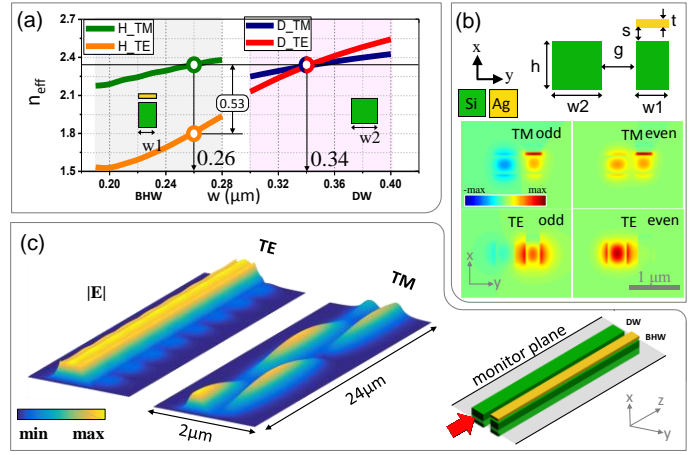

Figure 3. (a) The dependence of the effective indices of the BHW and DW on the width of waveguide (w). (b) Geometric details of the coupling region. ($w_1$, $w_2$, g, s, t) = (0.26, 0.34, 0.2, 0.05, 0.1) μm. The inserted figures are the odd and even modes for quasi-TE and -TM polarization states in the coupling region, respectively. (c) Electric field distributions for the quasi-TM mode and the quasi-TE mode in the monitor plane. The position of the monitor planes and the light launch position are shown at the right corner.

For the DW, the TE- and TM-like polarized modes have the same effective index ($n_{D-TM} = n_{D-TE} \approx 2.33$) when the width and height are equal ($w_2 = h = 0.34$ μm). By choosing $w_1 = 0.26$ μm, the BHW has the same effective index as the DW for the TM-like polarized light ($n_{D-TM} = n_{H-TM} \approx 2.33$). The odd and even modes for TE-like and TM-like polarization states in the coupling region are shown in Fig.3b. In case of perfect phase-matching, that is, $n_{D-TM} = n_{H-TM}$, the optical mode initially in the DW can make a complete transition to the BHW after propagating over a distance of $z_c$ ($z_c = \pi/[(n_{r\_e} - n_{r\_o})k_0]$), and *vice versa* (Fig. 3c), where $n_{r\_e}$ and $n_{r\_o}$ are the real parts of effective indices of the even and odd modes in the coupling region for the TM-like polarized light. However, due to the birefringence effect, the effective indices of the DW differs dramatically from the one of the BHW, since their difference is as large as 0.53 ($n_{D-TE} - n_{H-TE} \approx 2.33 - 1.80 = 0.53$). In this case of strong phase-mismatch, the TE-like polarized light initially in the DW can hardly couple into the adjacent BHW, as shown in Fig. 3c. Therefore, it is clear that the function of the polarization splitter is that the complete energy exchange is allowed for a quasi-TM mode while it is forbidden for the quasi-TE mode.

For the choice of the gap g = 200 nm between the DW and BHW, the effective indices of the even and odd modes for the quasi-TM mode is $n_{TM\_e} = 2.38718+0.00174273i$ and $n_{TM\_o} = 2.27618+0.00217367i$. Thus, the coupling length $z_c = \pi/[(n_{r\_e} - n_{r\_o})k_0] = \pi/[(2.38718-2.27618)*k_0] \approx 6.98$ μm. Considering that there are some undesired couplings between the S-bend parts, we choose the length of the straight part of BHW/DW as $L_2 = 5.4$ μm, a little shorter than $z_c$, so as to obtain a complete energy transition for the TM-like polarized light.

## 5. The whole structure simulation results

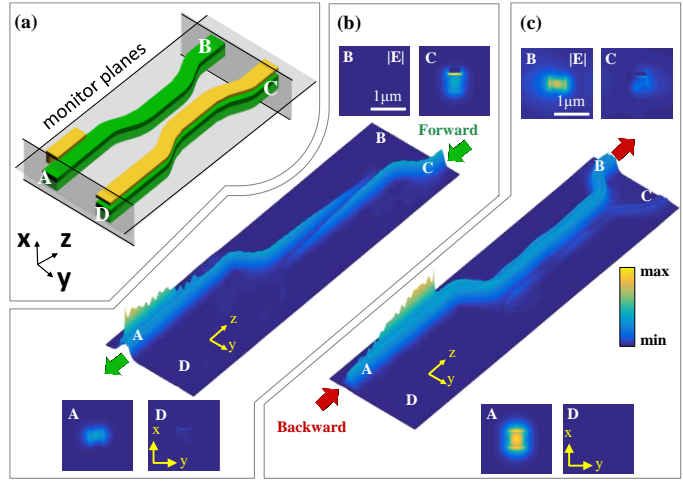

Figure 4. (a) The positions of the monitor planes. (b) Light propagates in the forward direction (quasi-TM mode launched in the port C), and the inserted figures are the electric field distribution detected from the monitor planes. (c) Light propagates in the backward direction (quasi-TM mode launched in the port A), and the inserted figures are the mode distribution detected from the monitor planes.

Figure 4a shows the monitor planes positions. It is now clear how this system functions. In the forward propagation, a quasi-TM mode launched in the input port C will firstly couple into the adjacent DW and then get converted into a quasi-TE mode after passing the RHW and finally reaches the output port A (Fig. 4b). In the backward propagation, the input quasi-TM mode at port A will firstly get converted into a quasi-TE mode via the RHW. However this newly quasi-TE mode cannot be coupled into the adjacent BHW due to the strong phase-mismatch, but instead, goes straightforward to the output port B (Fig. 4c).

Note that there will be a small proportion of light ends up going to an undesired port. For example, a little light goes to port D in the forward propagation. To evaluate the optical performance of this one-way operation, we defined the extinction ($E_{xt}$) as the ratio between the energy transmission detected in the desired port and that in the undesired port, that is $E_{xt} = 10*\log_{10}(P_d/P_u)$, assuming unit power entering the input port. For the forward propagation, the energy detected in port A (desired) and port D (undesired) are respectively $P_A$ = 50.515% and $P_D$ = 0.576%. Thus the forward extinction is calculated as $E_{xt\_F} \approx 19.43$ dB while the backward extinction calculated as $E_{xt\_B} \approx 11.82$ dB, provided that $P_B$ = 65.38% and $P_C$ = 4.301% in the backward propagation. The extinction ratio can be further improved by optimizing a few parameters, such as the gap (g) between two adjacent waveguides, the length of the straight part of BHW/DW ($L_2$), and the size of S-bend structure (a).

## 6. Operation bandwidth discussion

The above calculation results are based on the wavelength of 1.55μm. In order to find out the spectral dependence of the proposed device, we carried out simulations on the full-structure simulations from λ = 1.50 μm to λ = 1.60 μm for both forward and backward propagations. The results are show in Figure 5. For some logical computing circuits, it is desirable that the power in the desired port is much higher than the undesired port so that it is easier to make the judgement between two types of signals: "high" (1) and "low" (0). Therefore, a higher extinction is desirable in terms of logical level distinction (1 or 0). Figure 5 shows that the bandwidth for extinction > 10 dB for both forward and backward propagation is as large as 70 nm, from 1.51 μm to 1.58 μm. These results demonstrate that the device is a broadband device that has potential application value in the fields of optical logical circuits.

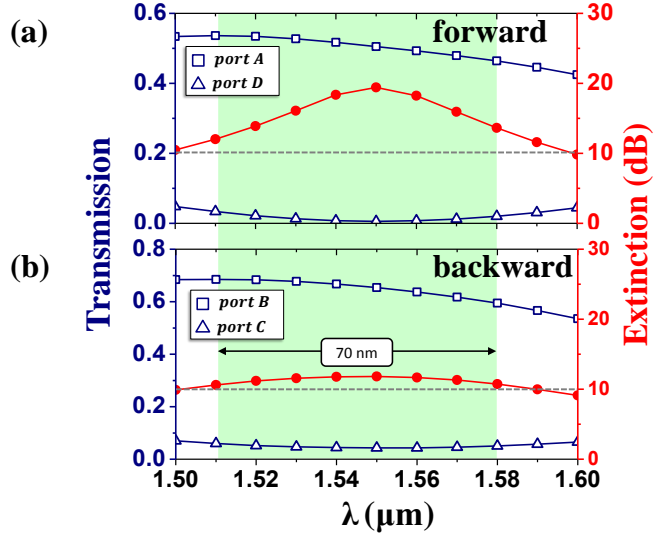

Figure 5. The transmission and extinction spectrum characteristics of forward (a) and (b) backward propagation directions. The blue curve indicates the output powers detected at the desired and undesired output ports. The red curve represents the dependence of the extinction on wavelength λ. The rectangle of light green color shows the bandwidth where the extinction ratio is greater than 10dB for both propagation directions.

## 7. Other functions and discussion

Interestingly, our proposed device can also be used as a polarization splitter-rotator [23, 24] by using the ports A, C and D (Figure 6a). When light propagates from port C to port A, the quasi-TM mode launched at port C is efficiently coupled to the adjacent DW at the directional coupler section and then the TM-like polarized light in the DW is converted to a TE-like polarized mode after passing through the RHW section. On the other hand, if one launches a TE-like polarization at the same input port C, the quasi-TE mode in the BHW cannot couple to the adjacent DW because of phase mismatch (see section 4 for details). Instead, it is restricted in the BHW and ends up going to port D (Figure 6b). This functionality is interesting. Not only can it separate different polarization state (TE- or TM-like polarization) in nanophotonic waveguide, but it can also guarantee that the polarization state (TE-like polarized) of every output port (port A or port D).

The proposed device is expected to have potential applications in optical logical computing. In conventionally logical computing, ideally, it is very desirable to have a full voltage (1) for a "high" state and zero voltage (0) for a "low" state. However, in reality, these perfect limits cannot be guaranteed due to stray voltage drops in the transistor circuitry. Similarly, in our proposed device, full optical power (1) and zero intensity (0) cannot be obtained at the output ports, because metallic structures in our device will inevitably cause some undesired optical losses since there is a weak coupling in the directional coupling section for quasi-TE modes, which will cause undesired coupling to the undesired port. In this case, we defined the output port with strong optical output power as logical high state (1) while the rest ports (with relative much lower optical output power) as logical low state (0). Thus, the functionality of the whole device can be described by a 8*8 matrix. Neglecting losses, its ideal form is as follows:

$$\begin{bmatrix} A_{TE\_out} \\ B_{TE\_out} \\ C_{TE\_out} \\ D_{TE\_out} \\ A_{TM\_out} \\ B_{TM\_out} \\ C_{TM\_out} \\ D_{TM\_out} \end{bmatrix} = \begin{bmatrix} 0 & 0 & 0 & 0 & 0 & 0 & 1 & 0 \\ 0 & 0 & 0 & 0 & 1 & 0 & 0 & 0 \\ 0 & 0 & 0 & 1 & 0 & 0 & 0 & 0 \\ 0 & 0 & 1 & 0 & 0 & 0 & 0 & 0 \\ 0 & 1 & 0 & 0 & 0 & 0 & 0 & 0 \\ 0 & 0 & 0 & 0 & 0 & 0 & 0 & 1 \\ 1 & 0 & 0 & 0 & 0 & 0 & 0 & 0 \\ 0 & 0 & 0 & 0 & 0 & 1 & 0 & 0 \end{bmatrix} \begin{bmatrix} A_{TE\_in} \\ B_{TE\_in} \\ C_{TE\_in} \\ D_{TE\_in} \\ A_{TM\_in} \\ B_{TM\_in} \\ C_{TM\_in} \\ D_{TM\_in} \end{bmatrix} \quad (3)$$

where A/B/C/D denote the physical ports while in/out represent the input and output ports, and TE/TM are the polarization states of input and output modes.

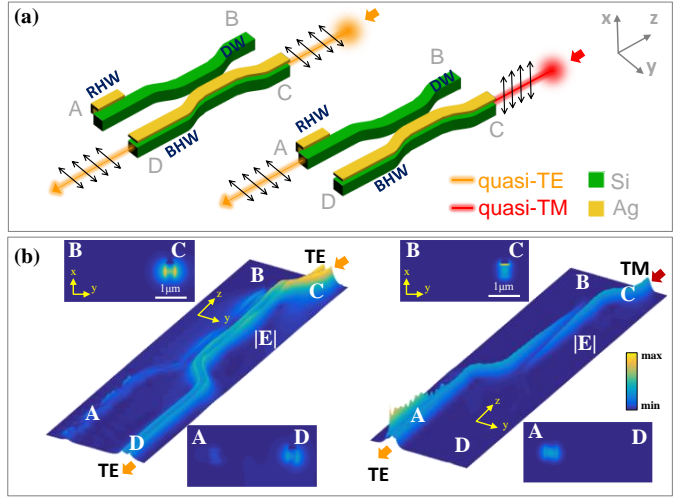

Figure 6. (a) Schematic illustration of the function of polarization splitter rotator. An input quasi-TE mode can travel from port C to port A without changing its original polarization state, while a quasi-TM mode launched in port C ends up going to port A with its original polarization state changing to TE-like polarization. (b) A quasi-TE mode or quasi-TM mode excited in the port C will respectively take different paths in the proposed device. The inserted figures are the mode distribution detected from the monitor planes. The positions of monitor planes are shown in Fig.4a.

Here we would like to emphasize that our device is conceptually different from chiral atom couplers [25-27], which can achieve unidirectional propagation of light by steering the handedness of transverse spin angular momentum (SAM) in the evanescent field of nanophotonic waveguides. The fundamental difference between those devices and our device is the underlying mechanism. They mainly take advantage of the transverse SAM in the evanescent field of nanophotonic waveguides to achieve chiral coupling, which is associate with the spin-orbit interaction of light [28]. They can achieve asymmetric propagation by changing the handedness of SAM of input light while we achieve the asymmetric propagation by breaking the polarization symmetry. Besides, they mainly handle the light field in the vicinity of waveguides while we mainly control the light inside nanophotonic waveguides. We believe that these two kind of mechanisms both have they own advantages and disadvantages, and they can find different applications in different fields. For example, the chiral coupler is much more compact compared with our device, but it suffers from low efficiency of chiral coupling, because only a small proportion of light can be coupled into the waveguides by atoms scattering [21]. On the other hand, our device is relative large in size, but it can handle all the light in nanophotonic waveguides.

Our device can be integrated into a complex photonic circuits as well. Two kind of waveguides are involved at the input and output ports, which are dielectric Si waveguide and hybrid plasmonic waveguide. In order to integrate this device into complex photonics circuits where there are different kind of waveguides, it requires some out-coupler structures to efficiently integrate it. For example, a tapper structure can be used as a coupler between the hybrid plasmonic waveguide and another kind Si waveguide [29]. Fortunately, this tapper coupler is also very compact with only several micrometers in size. Therefore, the use of out-coupler structure will not significant increase the compact dimensions of the device.

## 8. Conclusion

In summary, we proposed and numerically demonstrated a novel hybrid plasmonic structures which perform optical diode behavior. Unlike previous on-chip diodes based on asymmetric mode conversion, where multimode waveguides are involved, our diode is achieved in single mode waveguides via breaking the polarization symmetry. Moreover, our device have multiple functions as well, and it can be used as a new type of polarization splitter-rotator, who can separate different polarization state (TE- or TM-like polarization) of the input port while guaranteeing single polarization state (TE-like polarization) at the output ports. Besides, the proposed device is ultra-compact with a small footprint of $2.95 \times 14.18$ $\mu m^2$. The operation bandwidth of the device is as large as 70nm around $\lambda = 1550$ nm. Our work highlights the potential of hybrid plasmonic waveguide for the controlling of optical modes' polarization and propagation, and it may open up new avenues for efficient on-chip optical logical circuitry and optical information processing.

# Funding.

# Acknowledgment.

Our work Supported by the Innovation Project of Graduate School of South China Normal University